\documentclass[12pt]{article}

\usepackage{times}
\usepackage{graphicx}
\usepackage{times}

\newenvironment{sciabstract}{%
\begin{quote} \bf}
{\end{quote}}

\newcounter{lastnote}

\title{Imaging the Mott Insulator Shells\\ using Atomic Clock Shifts}

\author
{Gretchen K. Campbell,$^{\ast}$ Jongchul Mun, Micah Boyd, Patrick
Medley, \\Aaron E. Leanhardt,$^{1}$ Luis Marcassa,$^{2}$ David E.
Pritchard, and Wolfgang
Ketterle\\
\normalsize{MIT-Harvard Center for Ultracold Atoms, Research
Laboratory of Electronics}\\
\normalsize{ and Department of Physics,}\\
\normalsize{Massachusetts
Institute of Technology, Cambridge, MA 02139, USA}\\
\normalsize{$^{1}$ JILA, Boulder, CO 80309, USA}\\
\normalsize{$^{2}$ Instituto de Fisica de Sao Carlos - University of S\~ao Paulo, Brazil}\\
\\
\normalsize{$^\ast$To whom correspondence should be addressed;
E-mail:  gcampbel@mit.edu.} }

\date{}

\begin{document}


\baselineskip24pt


\maketitle


\begin{sciabstract}
Microwave spectroscopy was used to probe the superfluid-Mott
Insulator transition of a Bose-Einstein condensate in a 3D optical
lattice. Using density dependent transition frequency shifts we were
able to spectroscopically distinguish sites with different
occupation numbers, and to directly image sites with occupation
number $n=1$ to $n=5$ revealing the shell structure of the Mott
Insulator phase. We use this spectroscopy to determine the onsite
interaction and lifetime for individual shells.
\end{sciabstract}

The Mott insulator transition is a paradigm of condensed matter
physics.  It describes how electron correlations can lead to
insulating behavior even for partially filled conduction bands.
However, this behavior requires a commensurable ratio between
electrons and sites. If this condition on the density is not exactly
fulfilled, the system will still be conductive.  For neutral bosonic
particles, the equivalent phenomenon is the transition from a
superfluid to an insulator for commensurable densities.  In
inhomogeneous systems, as in atom traps, the  condition of
commensurability no longer applies: for sufficiently strong
interparticle interactions, it is predicted that the system should
separate into Mott insulator shells with different occupation
number, separated by thin superfluid layers
\cite{Jaksch98,Batrouni90,DeMarco05}.

The recent observation of the superfluid to Mott Insulator (MI)
transition with ultracold atoms \cite{Greiner02} has stimulated a
large number of theoretical and experimental studies [see
\cite{Bloch05} and references therein]. Atomic systems allow for a
full range of control of the experimental parameters including
tunability of the interactions and defect-free preparation making
them attractive systems for studying condensed matter phenomena. The
Mott insulator phase in ultracold atoms has been characterized by
studies of coherence, the excitation spectrum, and noise
correlations \cite{Greiner02,Schori04,Folling05}. Recently, using
spin-changing collisions, lattice sites with two atoms were
selectively addressed and the suppression of number fluctuations was
observed \cite{Gerbier06}.

The most dramatic new feature of the MI phase in ultracold atoms is
the layered structure of the Mott shells, however until now this
feature has not been directly observed.  In this paper, we combine
atoms in the Mott insulator phase with the high resolution
spectroscopy used for atomic clocks, and use density dependent
transition frequency shifts to spectroscopically resolve shells with
occupancies from $n$ = 1 to 5 and directly image their spatial
distributions.

Bosons with repulsive interactions in an optical lattice are
accurately described by the Hamiltonian \cite{Fisher89,Jaksch98},
\begin{equation}
\hat{H}=-J\sum_{\langle i,j\rangle} \hat{a}_\imath^\dag
a_\jmath+\frac{1}{2}U\sum_i \hat{n}_i(\hat{n}_i-1)+{\sum_i
(\epsilon_i-\mu)\hat{n}_i},
\end{equation}
where the first two terms are the usual Hamiltonian for the
Bose-Hubbard model and the last term adds in the external trapping
potential, and where $J$ is the tunneling term between nearest
neighbors, $\hat{a}_i^\dag$ and $\hat{a}_i$ are the boson creation
and destruction operators at a given lattice site,
$U=(4\pi\hbar^2a/m)\int|w(x)|^4d^3x$ is the repulsive onsite
interaction, where $m$ is the atomic mass, $a$ is the s-wave
scattering length and $w(x)$ is the single particle Wannier function
localized to the $i^{th}$ lattice site, and
$\hat{n}_i=\hat{a}_i^\dag\hat{a}_i$ is the number operator for
bosons at site $i$. The last term in the hamiltonian is due to the
external trapping confinement of the atoms, where
$\epsilon_i=V_{ext}(r_i)$ is the energy offset at the $i$th site due
to the external confinement and $\mu$ is the chemical potential.

The behavior of this system is determined by the ratio of $J/U$. For
low lattice depths, the ratio is large and the system is superfluid.
For larger lattice depths, the repulsive onsite energy begins to
dominate, and the system undergoes a quantum phase transition to a
MI phase. For deep lattices the atoms are localized to individual
lattice sites with integer filling factor $n$.  This filling factor
varies locally depending on the local chemical potential
$\mu_i=\mu-\epsilon_i$ as
\begin{equation}
 n=Mod(\mu_i/U),
 \end{equation}
and decreases from the center to the edge of the trap.

To prepare the atoms in the Mott insulating phase, a $^{87}$Rb
Bose-Einstein condensate was first created in the the
$|F=1,m_F=-1\rangle$ state using a combination of a Ioffe-Pritchard
magnetic trap and an optical dipole trap.  The optical trap was
oriented perpendicular to the long axis of the magnetic trap,
creating a more isotropic trapping potential which was better
matched to the optical lattice. The laser beam for the optical trap
had a $1/e^2$ waist of $\approx$ 70 $\mu$m, and was retro-reflected.
However, the polarization of the retro-reflected  beam was rotated
such that the interference between the two beams had minimal
contrast. The resulting trap had radial (axial) trap frequencies of
$\omega$ = 2$\pi\times$ 70 (20) Hz, where the axial direction is now
parallel to the optical trap. A 3D optical lattice was created by
adding two additional retro-reflected laser beams derived from the
same laser at $\lambda$ = 1064 nm. The lattice was adiabatically
ramped up by rotating the polarization of the retro-reflected
optical trapping beam to increase the interference contrast along
that axis and by increasing the laser power in the other two axes.
The lattice depth was increased using an exponential ramp with a 40
ms time constant. All three beams were linearly polarized orthogonal
to each other and had different frequency detunings generated using
acousto-optic modulators. The lattice depth was up to 40 $E_{rec}$,
where $E_{rec}=\hbar^2 k^2/(2m)$ and $k=2\pi/\lambda$ is the
wavevector of the lattice. At 40 $E_{rec}$ the lattice trap
frequency at each site was $\omega_{lat}$ = 2$\pi$ $\times$ 25 kHz,
and the external trap frequencies increased to $\omega$ =
2$\pi\times$ 110 (30) Hz in the radial (axial) direction.

Zeeman shifts and broadening of the clock transition from the $F=1$
to the $F=2$ state were avoided by using a two photon transition
between the $|1,-1\rangle$ and the $|2,1\rangle$ state, where at a
magnetic bias field of $\sim3.23$ G both states have the same first
order Zeeman shift \cite{Harber02}. The two photon pulse was
composed of one microwave photon at a fixed frequency of 6.83 GHz,
and one rf photon at a frequency of around 1.67 MHz. The pulse had a
duration of 100 ms, and when on resonance the fraction of atoms
transferred to the $|2,1\rangle$ state was less than 20\%. After the
pulse, atoms in the $|2,1\rangle$ state were  selectively detected
with absorption imaging using light resonant with the $5^2S_{1/2}
|2,1\rangle \rightarrow 5^2P_{3/2}|3,1\rangle$ transition. For
observing the spatial distribution of the Mott shells, the atoms
were imaged in trap. For recording spectra, the atoms were released
from the trap and imaged after 3 ms of ballistic expansion in order
to reduce the column density

When the two photon spectroscopy is performed on a trapped
condensate without a lattice, the atoms transferred to the
$|2,1\rangle$ state have a slightly different mean field energy due
to the difference between the $a_{21}$ and $a_{11}$ scattering
lengths, where $a_{21}$ is the scattering length between two atoms
in states $|2,1\rangle$ and $|1,-1\rangle$, and $a_{11}$ is the
scattering length between two atoms in the $|1,-1\rangle$ state.
This difference in the scattering lengths leads to a density
dependent shift to the resonance frequency $\Delta\nu\propto\rho
(a_{21}-a_{11})$, where $\rho$ is the condensate density
\cite{Harber02}. This collisional shift is commonly referred to as
the `clock' shift \cite{Gibble93} due to its importance in atomic
clocks, where cold collisions currently limit the accuracy
\cite{Fertig00,Sortais01}. When performed on a condensate with peak
density $\rho_0$ in a harmonic trap, in the limit of weak
excitation, the line shape for the 2-photon resonance is given by
\cite{Stenger99}:
\begin{equation}
I(\nu)=\frac{15 h (\nu-\nu_0)}{4 \rho_0\Delta
E}\sqrt{1-\frac{h(\nu-\nu_0)}{\rho_0\Delta E}},
\end{equation}
with the mean field energy difference
\begin{equation}
\Delta E=\frac{h^2}{\pi m}\left(a_{21}-a_{11}\right).
\end{equation}
In the case of $^{87}$Rb, $a_{21}$($a_{11}$) = 5.19 (5.32) nm
\cite{VanKempen02}. Both the frequency shift and the linewidth
increase with the condensate density. As the lattice is ramped on,
the peak density of the condensate in a given lattice site increases
as
\begin{equation}
\rho_0(r)=\left(\mu-\frac{1}{2}m\omega_{trap}^2r^2\right)1/U,
\end{equation}
where $\omega_{trap}$ is the external trap frequency for the
combined magnetic and optical trap, and using the Thomas Fermi
approximation $\mu$, the chemical potential is given by
\begin{equation}
\mu=\left(\frac{15}{16}\frac{(\lambda/2)^3m^{3/2}NU\omega_{trap}^3}{\sqrt{2}\pi}\right)^{2/5},
\end{equation}
where $N$ is the total atom number. For low lattice depths the
system is still a superfluid, delocalized over the entire lattice.
However, the two-photon resonance line is shifted and broadened due
to the increased density, with the center of the resonance at $\nu$
= $\nu_0+2 \rho_0\Delta E/3h$. For deep lattices in the MI regime
the repulsive onsite interaction dominates, number fluctuations are
suppressed, and each lattice site has a sharp resonance frequency
determined by the occupation number in the site. The separation
between the resonance frequencies for the $n$ and $n-1$ MI phases is
given by
\begin{equation}
\delta\nu=\frac{U}{h}\left(a_{21}-a_{11}\right)/a_{11}.
\end{equation}
The linewidth of the resonances is no longer broadened by the
inhomogeneous density and should be limited only by the bandwidth of
the two photon pulse.

\begin{figure}[ht!!!]
\centering{
\includegraphics[width=5.0in,]{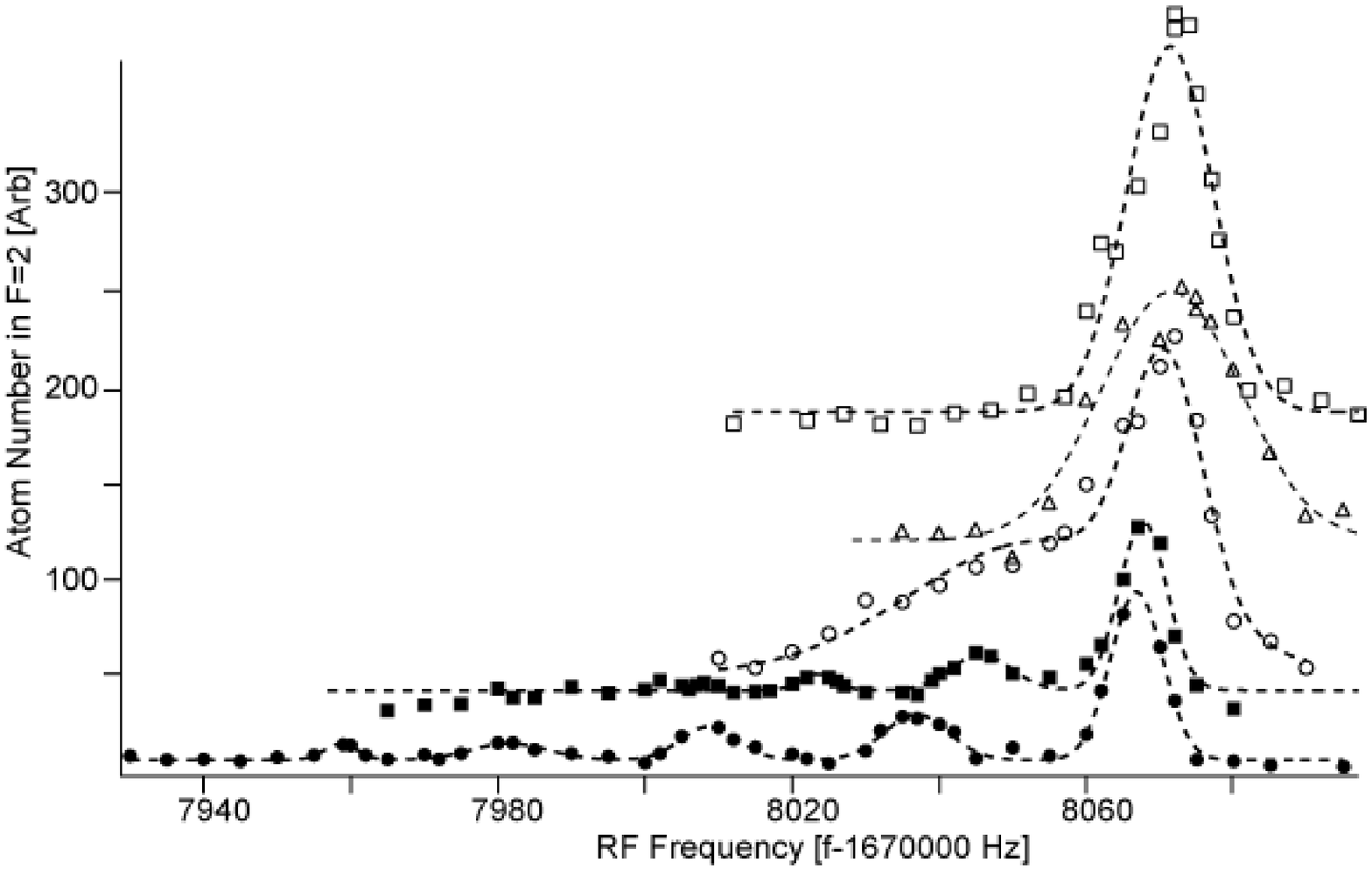}
\caption{\label{fig:figtwo} }}
\end{figure}

Fig.~\ref{fig:figtwo} shows the transition from a broadened
resonance to several sharp lines as the lattice depth was increased.
At a lattice depth of V = 5 $E_{rec}$, the line was broadened and
the line center was shifted slightly due to the increased density.
At V = 10 $E_{rec}$, the line was shifted and broadened further, and
in addition the line shape became asymmetric as the atom number in
lattice sites with small occupation was squeezed. For deeper lattice
depths the system underwent a phase transition to a MI phase and
discrete peaks appeared corresponding to MI phases with different
filling factors, and for V = 35$E_{rec}$, MI phases with occupancies
of up to 5 were observed.

\begin{figure}
\centering{
\includegraphics[hiresbb=true,width=3.7in]{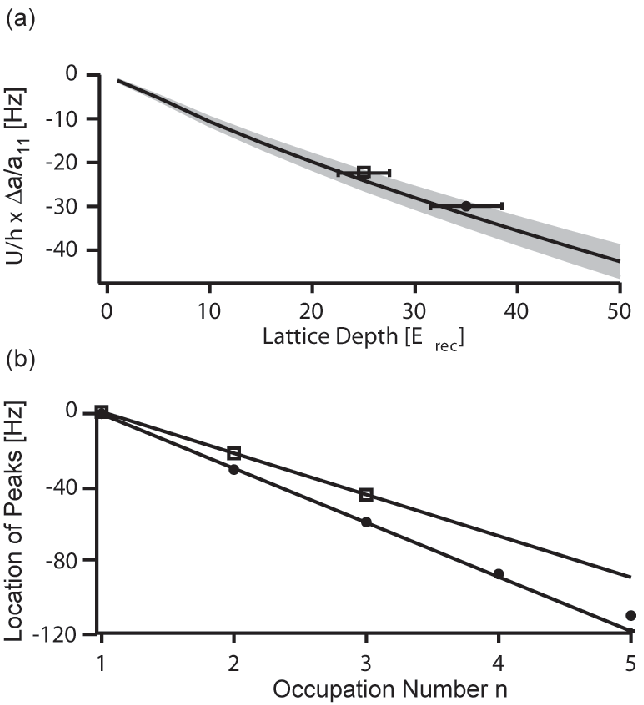}
\caption{\label{fig:figfour}}}
\end{figure}

When the lattice depth was increased inside the MI regime (from V =
25 $E_{rec}$ to 35 $E_{rec}$), the separation between the resonance
peaks increased presumably due to the larger onsite interaction
energy as the lattice trap was increased. As given in Eq. (7), the
separation between the peaks provides a direct measurement of the
onsite interaction energy $U$. As shown in Fig.~\ref{fig:figfour}a,
our results are in good agreement with calculated values of $U$.
Fig.~\ref{fig:figfour}b shows that although the separation between
the $n$ = 1, 2 and 3 peaks is approximately constant, for higher
filling factors the separation between the peaks decreases; the
effective onsite interaction energy becomes smaller for higher
filling factors. This shows that for low occupation numbers the
atoms occupy the ground state wavefunction of the lattice site,
whereas for larger occupation numbers, the repulsive onsite
interaction causes the wavefunction to spread out lowering the
interaction energy. From a variational calculation of the
wavefunction similar to \cite{Baym96}, we find the onsite energy for
the $n$ = 5 shell should be $\approx$ 20\% smaller than that for the
$n$ = 1 shell, is in agreement with the measured value shown in
Fig.~\ref{fig:figfour}b.

\begin{figure}
\centering{
\includegraphics[hiresbb=true,width=4.0in]{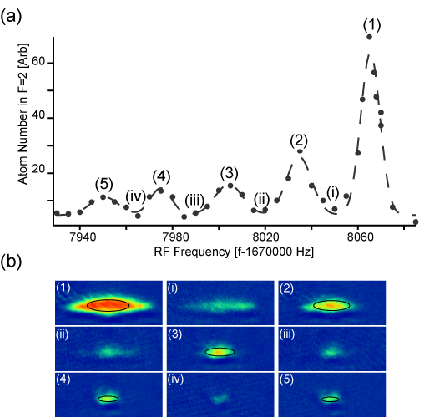}
\caption{\label{fig:figthree} }}
\end{figure}

The peaks for the different occupation numbers were spectrally well
separated.  Therefore, on resonance, only atoms from a single shell
were transferred to the $|2,1\rangle$ state. An image of these atoms
(without any time-of-flight) shows the spatial distribution of this
shell. Fig.~\ref{fig:figthree}b, shows absorption images for $n$ = 1
- 5 shells.  As predicted \cite{Jaksch98}, the $n$ = 1 MI phase
appears near the outer edge of the cloud.  For larger $n$, the
radius of the shell decreases, and the $n$ = 5 sites form a core in
the center of the cloud. The expected radius for each shell was
obtained from Eq. 2 using the measured values for the onsite
interaction.   The observed radii were in good agreement, except for
the $n$ = 1 shell which may have been affected by anharmonicities in
the external trap. Absorption images taken with rf frequencies
between the peaks show a small signal, which may reflect the
predicted thin superfluid layers between the insulating shells,
however this needs be studied further. The expected absorption image
of a shell should show a column density with a flat distribution in
the center and raised edges. However due to limitations (resolution
and residual fringes) in our imaging system, these edges were not
resolved.

\begin{figure}
\centering{
\includegraphics[hiresbb=true,width=4.0in]{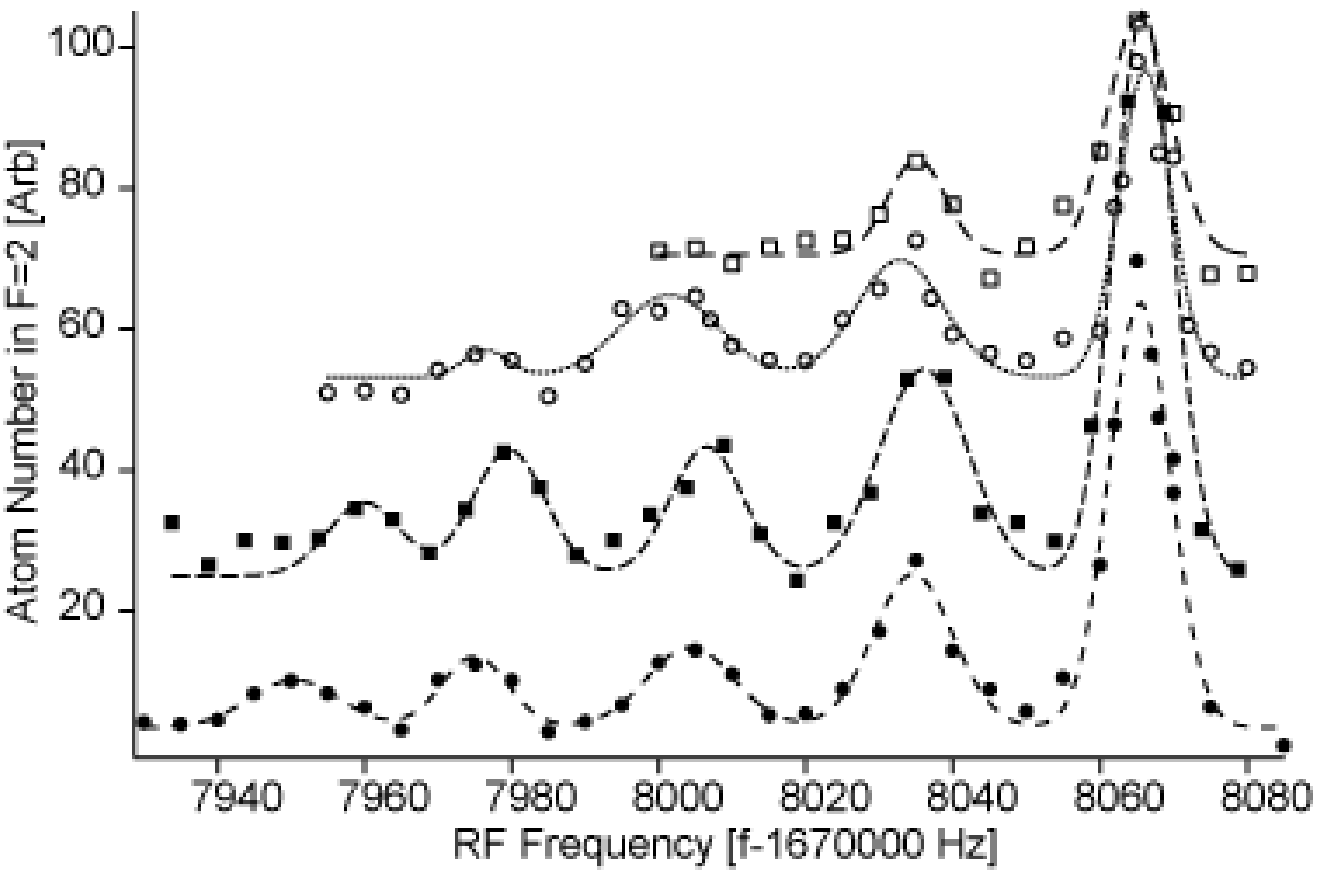}
\caption{\label{fig:figfive} }}
\end{figure}

Since we were able to address the different MI phases separately,
the lifetime for each shell could be determined.  For this, the
atoms were first held in the lattice for a variable time $\tau$
before applying the 100 ms two-photon pulse. For the $n$ = 1 MI
phase, ignoring technical noise, the lifetime should only be limited
by spontaneous scattering from the lattice beams.  Even for the
deepest lattices, the spontaneous scattering rate is $<10^{-2}$ Hz.
For the $n$ = 2 MI phase the lifetime is limited by dipolar
relaxation, which for $^{87}$Rb is slow with a rate $<10^{-2}$ Hz.
For sites with $n\geq$ 3 the lifetime is limited by 3-body
recombination with a rate $\gamma n(n-1)(n-2)$ \cite{Jack05} with
$\gamma$ = 0.026 Hz for our parameters.  This gives 3-body lifetimes
of of $\tau_{3B}$ = 6.2 s, 1.6 s, and 0.6 s for the $n$ = 3, 4,  and
5 MI phases respectively. This calculation of $\gamma$ assumes for
the density distribution the ground state of the harmonic oscillator
potential, so for higher filling factors the actual lifetime could
be higher. In Fig.~\ref{fig:figfive}, we show relative populations
as a function of the hold time and derive lifetimes as $\tau\approx$
1 s, 0.5 s, and 0.2 s for the $n$ = 3, 4, and 5 MI phases, this is
shorter than predicted which is possibly due to secondary
collisions. For $n$ = 1 and 2, lifetimes of over 5 s were observed.

In this work we have used an atomic clock transition to characterize
the superfluid - Mott insulator transition.  This demonstrates the
power of applying precision methods of atomic physics to problems of
condensed matter physics. Exploiting atomic clock shifts, we could
distinguish sites according to their occupation number.  This
allowed us to directly image the Mott insulator shells with filling
factors $n$ = 1 - 5, and to determined the onsite interaction and
lifetime as a function of $n$.

In the future, this method can be used to measure the number
statistics as the system undergoes the phase transition.  One would
expect that the spectral peaks for higher occupation number become
pronounced only at higher lattice depth --- an indication of this
can be seen already in Fig. 1.  For low lattice depths, the
tunneling rate is still high, but one can suddenly increase the
lattice depth and freeze in populations \cite{Greiner02b}, which can
then be probed with high resolution spectroscopy.  Fluctuations in
the atom number could identify the superfluid layers between the
Mott shells. In addition, by applying a magnetic gradient across the
lattice, tomographic slices could be selected, combining full 3D
resolution with spectral resolution of the site occupancy. The
addressability of individual shells  could be used to create systems
with only selected occupation numbers (e.g. by removing atoms in
other shells). Such a preparation could be important for the
implementation of quantum gates for which homogenous filling is
desirable. For atoms other than rubidium, atomic clock shifts are
much larger, e.g. for sodium by a factor of 30. Therefore, it should
be easier to resolve the Mott insulator shells, unless the
collisional lifetime of the upper state of the clock transition sets
a severe limit to the pulse duration \cite{thanks}.

\bibliographystyle{Science}

\begin{thebibliography}{10}
\bibitem{thanks}
This work was supported by NSF.
\end{thebibliography}

\newpage

\noindent {\bf Fig. 1.}Two-photon spectroscopy across the superfluid
to Mott insulator transition. Spectra for 3D lattice depths of 0
$E_{rec}$ (open squares), 5 $E_{rec}$ (open diamonds), 10 $E_{rec}$
(open circles), 25 $E_{rec}$ (solid squares), and 35 $E_{rec}$
(solid circles) are shown. The spectra are offset for clarity. The
shift in the center of the $n$=1 peak as the lattice depth is
increased is due to the differential AC Stark shift from
the lattice. The dotted lines show gaussian fits of the peaks.\\
\newline
\newpage

\noindent {\bf Fig. 2.} Probing the onsite interaction energy. (a)
The separation between the $n$ = 1 and $n$ =2 peaks is shown for
lattice depths of V = 25 $E_{rec}$ (squares) and V = 35 $E_{rec}$
(circles). As the lattice depth was increased the separation
increased from 22(1) Hz to 30(1) Hz. The shaded area gives the
expected value determined from a band structure calculation,
including the uncertainty in the scattering lengths. The uncertainty
in the measured separation is indicated by the size of the points.
(b) Location of resonances for all MI phases relative to the $n$ = 1
phase for V = 25 $E_{rec}$ and V = 35 $E_{rec}$. For low site
occupation ($n$ = 1, 2, 3), the separation between the resonances is
approximately constant, implying constant on-site interaction energy
U.  For V = 35 $E_{rec}$, the separation between the $n$ = 4 and 5
peaks was 22(2) Hz, a 27\% decrease from the 30(1) Hz separation
between the $n$ = 1 and 2 peaks. The slope of the
lines is fit to the separation between the $n$ = 1 and 2 peaks.\\
\newline

\noindent {\bf Fig. 3.} Imaging the shell structure of the Mott
insulator. (a) Spectrum of the Mott insulator at V = 35 $E_{rec}$.
(b) Absorption images for for decreasing rf frequencies. Images $1$,
$2$, $3$, $4$ and $5$ were taken on resonance with the peaks shown
in (a) and display the spatial distribution of the $n$ =1-5 shells.
The solid lines shows the predicted contours of the shells.
Absorption images taken for rf frequencies between the peaks (images
$i$, $ii$ ,$iii$ and $iv$) show a much smaller signal. The field of
view was 185 $\mu$m $\times$ 80 $\mu$m.\\
\newline
\newpage

\noindent {\bf Fig. 4.} Lifetime of individual MI shells. The
lifetime for each MI phases can be measured independently by adding
a hold time before applying the two photon pulse. Spectra are shown
for hold times of 0 ms (filled circles), 100 ms (filled squares),
400 ms (open circles), 2000 ms (open squares). The lattice depth was
V = 35 $E_{rec}$ except for the 100 ms hold time, for which it was
V=34 $E_{rec}$. The lines show gaussian fits to the peaks, and the
spectra were offset for clarity.\\
\newline

\end{document}